\documentclass[11pt]{article}
\usepackage{amssymb,amsmath,cite,graphicx,geometry,mathrsfs}
\catcode`\@=11

\@addtoreset{equation}{section}
\def\theequation{\arabic{section}.\arabic{equation}}
     
\catcode`\@=11
\def\thesection{\arabic{section}}

\def\appendix{\setcounter{section}{0}
        \def\thesection{Appendix \Alph{section}}
        \def\theequation{\Alph{section}.\arabic{equation}}}
\def\section{\@startsection{section}{1}{\z@}{3.5ex plus 1ex minus
   .2ex}{2.3ex plus .2ex}{\large\bf}}
   
\long\def\@makefntext#1{\parindent 0cm\noindent
\hbox to 1em{\hss$^{\@thefnmark}$}#1}

\newcommand{\captionfonts}{\small}
\makeatletter  
\long\def\@makecaption#1#2{%
  \vskip\abovecaptionskip
  \sbox\@tempboxa{{\captionfonts #1: #2}}%
  \ifdim \wd\@tempboxa >\hsize
    {\captionfonts #1: #2\par}
  \else
    \hbox to\hsize{\hfil\box\@tempboxa\hfil}%
  \fi
  \vskip\belowcaptionskip}
\makeatother   

\newcommand{\tq}{\triangleq}

\begin{document}
\begin{titlepage}
\vspace{.5in}
\begin{flushright}
September 2019\\  
\end{flushright}
\vspace{.5in}
\begin{center}
{\Large\bf
 Near-Horizon BMS Symmetry,\\[1.1ex]
 Dimensional Reduction, 
and Black Hole Entropy }

\vspace{.4in}
{S.~C{\sc arlip}\footnote{\it email: carlip@physics.ucdavis.edu}\\
       {\small\it Department of Physics}\\
       {\small\it University of California}\\
       {\small\it Davis, CA 95616}\\{\small\it USA}}
\end{center}

\vspace{.5in}
\begin{center}
{\large\bf Abstract}
\end{center}
\begin{center}
\begin{minipage}{4.75in}
{\small 
In an earlier short paper [Phys.\ Rev.\ Lett.\ 120 (2018) 101301, arXiv:1702.04439],
I argued that the horizon-preserving diffeomorphisms of a generic black hole are enhanced 
to a larger BMS${}_3$ symmetry, which is powerful enough to determine the Bekenstein-Hawking 
entropy.  Here I provide details and extensions of that argument, including a 
loosening of horizon boundary conditions and a more thorough treatment of dimensional 
reduction and meaning of a ``near-horizon symmetry.''}
\end{minipage}
\end{center}
\end{titlepage}
\addtocounter{footnote}{-1}

\section{Introduction}

The discovery by Bekenstein \cite{Bekensteinx} and Hawking \cite{Hawkingx} that
black holes are thermodynamic objects has led to a host of fascinating puzzles,
from the information loss problem to the question of what microscopic states are
responsible for black hole entropy.  Here, I focus on one particular puzzle,
the ``problem of universality'' of black hole entropy.  This paper is an expanded version 
of a short article published in 2018 \cite{Carlipxa}; here I discuss details and extend
some of the results.

The universality of black hole entropy has two aspects, probably related but logically
distinct.  The first comes from the simple form of the Bekenstein-Hawking entropy, 
\begin{align}
S_{\hbox{\tiny\it BH}} = \frac{A_{\hbox{\tiny\it hor}}}{4G\hbar} \, ,
\label{intro1}
\end{align}
where  $A_{\hbox{\tiny\it hor}}$ is the horizon area.  If this were merely a property
of, say, uncharged static black holes, it would tell us something important about
the Schwarzschild solution.  But it is more.  The same area law, with the same 
coefficient, holds for black holes with any charges, any spins, in any dimensions;
it holds for black strings, black rings, black branes, and black Saturns (black holes
encircled by black rings); it remains true for ``dirty black holes'' whose
horizons are distorted by nearby matter.  The only known way to change the entropy
(\ref{intro1}) is to change the Einstein-Hilbert action, and even then the correction
will be another universal term \cite{Iyer}.   Black hole entropy is not, it seems, a property of
specific solutions, but rather a generic characteristic of horizons.

The second aspect of universality emerges when one attempts to identify the 
microscopic states responsible for this entropy.  We do not yet have a 
full description of those states; that would presumably require a complete quantum 
theory of gravity.  We do, however, have an assortment of research programs working 
toward the quantization of gravity, which allow partial computations of black hole entropy.  
In string theory, black
hole entropy can be calculated from properties of weakly coupled strings and branes, 
from the AdS/CFT correspondence, and (probably) from an enumeration of horizonless
``fuzzball'' configurations.  In loop quantum gravity, entropy can be calculated from
a horizon Chern-Simons theory, from an analysis  of spin network reconnections in
the interior, and from conformal field theory at ``punctures'' of the horizon.  In induced 
gravity---an approach in which the Einstein-Hilbert action is obtained by integrating out 
``heavy'' fields in the path integral---entropy can be calculated from the
properties of the heavy fields.  In semiclassical gravity, entropy can be calculated from
either a single instanton approximation or pair production.  And, of course,
entropy can be calculated using Hawking's original approach, which involved only quantum
field theory in a fixed black hole background.

None of these methods is complete.  String theory calculations, for instance, are 
cleanest for near-extremal black holes, while loop quantum gravity calculations
may depend on a new universal constant, the Barbero-Immirzi parameter.
But although they describe very different microstates, each of these methods, within 
its range of validity, reproduces the standard Bekenstein-Hawking entropy.  (For a 
review and further references, see \cite{Carlipz}.)

One might worry about a 
selection effect here: perhaps models that give the ``wrong'' entropy are less likely to 
be published.  But even the elegant analysis of BPS black holes in string theory 
\cite{Vafa}, the first really successful microscopic calculation of black hole entropy,
illustrates the problem.  Given a spacetime dimension and a set of charges and 
spins, one can calculate the entropy of a gas of strings and branes at weak coupling;
separately calculate the horizon area of a black hole at strong coupling; and compare 
the results.   But although the final answer always matches the Bekenstein-Hawking 
area law (\ref{intro1}), each new choice of dimension, spins, and charges requires a 
new computation.  Some underlying structure is clearly missing.

A first guess for this deeper structure is that the relevant degrees of freedom live on 
the horizon \cite{Bekensteinx}.  But this is not enough: while it could explain an 
area law for black hole entropy, there is no obvious reason why the coefficient $1/4$ 
should be universal.  
An elaboration of this idea, first suggested (I believe) in \cite{Carlip1}, is that the 
entropy is governed by a horizon symmetry.  This is, of course, a very strong requirement:
symmetries can place some restrictions on the density of states, but they are 
rarely strong enough to actually determine the entropy.  But we know one  symmetry that 
has the same kind of universal properties we see in black hole entropy.  As Cardy first 
showed in 1986 \cite{Cardy,Cardyb}, two-dimensional 
conformal symmetry is so restrictive that it completely fixes the asymptotic density 
of states in terms of a few parameters, independent of any of the fine details of the theory.
More recently, it has been shown that a related symmetry, that of the three-dimensional 
Bondi-Metzner-Sachs group (BMS${}_3$), exhibits the same universality \cite{Bagchi}.  
The possibility of a connection with black hole entropy has obvious appeal.

This connection was first confirmed for the (2+1)-dimensional BTZ black hole in 1998
 \cite{Strominger,BSS}.  Attempts to extend those results to higher dimensions 
soon followed \cite{Carlip2,Solodukhin}.  These efforts, which typically involve a 
search for a suitable two-dimensional group of horizon symmetries,  have had 
significant success; see \cite{Carlip3} for a review.  But they have been plagued by 
several problems:
\begin{itemize}
\item The symmetries are typically located either at infinity or on a timelike ``stretched 
horizon'' just outside the actual horizon (although with occasional exceptions 
\cite{Paddy}).  The physics at infinity is extremely 
powerful, especially for asymptotically anti-de Sitter spaces.  Indeed, the BTZ black hole
calculations were among the first examples of the now famous AdS/CFT correspondence.  But
the symmetries alone are not enough; by themselves, for instance, they cannot distinguish a 
black hole from a star with the same mass.  

The stretched horizon more directly captures the local properties of the black hole.  But 
the definition of the stretched horizon is not unique, and different limits can lead to
different entropies \cite{Silva,Carlip4}.  Moreover, while the entropy has a well-defined 
limit at the horizon, other parameters in the symmetry algebra typically blow up at
the horizon\cite{Carlip5,Dreyer,Koga} (again with occasional exceptions \cite{Dreyer2}).
\item The  approach fails in what should be the simplest case, two-dimensional dilaton 
gravity, where the zero-dimensional boundary of a Cauchy surface simple doesn't
have ``room'' for the required central term in the conformal algebra.  There are ad hoc 
fixes---lifting the theory to three dimensions \cite{Navarro} or artificially introducing an 
integral over time \cite{Mignemi}---but none of them is very convincing.  
\item In higher dimensions, the relevant symmetries are those of the ``$r$--$t$ plane'' 
picked out by the horizon.  But to obtain a well-behaved symmetry algebra, one must 
introduce an extra ad hoc dependence on angles, with no clear physical 
justification.  This is especially problematic for the Schwarzschild black hole, for which
the angular dependence breaks spherical symmetry in a manner that seems quite
arbitrary.
\end{itemize}

Here (and in in a shorter form in \cite{Carlipxa}) I describe an approach that avoids 
these problems.  A basic limitation of past work, I argue, was the attempt to force
the horizon symmetry into the form of a two-dimensional conformal symmetry.  This 
was an understandable choice: until quite recently, this was the only symmetry known 
to be powerful enough to determine the asymptotic density of states.  But with the 
discovery that BMS${}_3$  symmetry also has this universal property, the possibilities 
have expanded.

Starting with the intuitive idea that the relevant symmetries should lie in the ``$r$--$t$ 
plane'' picked out by the null generators of the horizon \cite{Carlip2,Carlip5,Dreyer2}, I 
first show how to reduce the problem to an effective two-dimensional model.  For spherically
symmetric black holes, such a dimensional reduction already appeared in some
of the earliest work on horizon symmetries \cite{Solodukhin,Giacomini}, but I demonstrate 
that the relevant near-horizon properties are far more general.  I next establish 
that the obvious horizon symmetries, the horizon-preserving diffeomorphisms, 
are enhanced by a particular shift invariance, as anticipated in \cite{Carlip6,Kang}.  This 
new symmetry may be viewed as a generalization of the global conformal symmetry found 
by Wall for horizon quantum field theory on a fixed background \cite{Wall}; it is exact  at the 
horizon and, in a sense I explain, it can be made arbitrarily close to exact near 
the horizon.    Using covariant phase space methods \cite{Carlip5,Barnich}, I show that 
the generators of these symmetries can be expressed as integrals along the horizon,
 with no need to go to a ``stretched horizon.''  Finally, I confirm that the resulting generators
satisfy a centrally extended BMS${}_3$  algebra that determines the correct Bekenstein-Hawking
entropy.

\section{BMS${}_3$ symmetry \label{secBMS}}

Let us start with a brief review of BMS${}_3$ symmetry, along with a discussion of the
perhaps puzzling question of how a classical symmetry can determine the number of
quantum states.

The BMS${}_3$ algebra is described by two sets of generators $L_n$ and $M_n$
($n\in \mathbb{Z}$) with Poisson brackets
\begin{align}
&i\left\{L_m,L_n\right\} = (m-n)L_{m+n} \, , \nonumber\\
&i\left\{M_m,M_n\right\} = 0 \, , \label{xxe1}\\
&i\left\{L_m,M_n\right\} = (m-n)M_{m+n} + c_{\scriptscriptstyle LM}m(m^2-1)\delta_{m+n,0} \, ,
\nonumber
\end{align}
where $c_{\scriptscriptstyle LM}$ is a classical central charge.\footnote{Other central
elements can also be added, but only $c_{\scriptscriptstyle LM}$ is relevant in the
present context.}  This algebra can be obtained as a contraction of the usual two-dimensional 
conformal (Virasoro) algebra, and is also isomorphic to the two-dimensional 
Galilean Conformal Algebra \cite{GCA}.  While the BMS${}_3$ algebra is not as thoroughly
studied as the conformal algebra, a fair amount is understood about its properties and
representations \cite{Oblak,Oblakb,Oblakc,Bagchix,Bagchiy}.

We shall see below that this algebra describes the classical horizon symmetry of   
generic black hole.  Let us assume that the same symmetry, perhaps 
deformed, is realized in the quantum theory.  We make the the usual substitutions 
\begin{align}
&\{{\scriptstyle\bullet},{\scriptstyle\bullet}\} \rightarrow
   \frac{1}{i\hbar}[{\scriptstyle\bullet},{\scriptstyle\bullet}] \, ,\nonumber\\
&\frac{1}{\hbar}L\rightarrow{\hat L} \, , \quad 
\frac{1}{\hbar}M\rightarrow{\hat M} \, , \quad\frac{1}{\hbar}c\rightarrow{\hat c}  \, ,
\label{xye1}
\end{align}
where the factors of $\hbar$ in $\hat L$ and $\hat M$ ensure that the operators are 
dimensionless.  (It is not always discussed explicitly in quantum mechanics textbooks, 
but the same substitution is used to go from the Poisson algebra of 
angular momentum to the Lie algebra of rotations.)   
We thus obtain a quantum operator algebra
\begin{align}
&[{\hat L}_m,{\hat L}_n]= (m-n){\hat L}_{m+n} \, , \nonumber\\
&[{\hat M}_m,{\hat M}_n] = 0 \, , \label{xxe2}\\
&[{\hat L}_m,{\hat M}_n] 
   = (m-n){\hat M}_{m+n} +  {\hat c}_{\scriptscriptstyle LM} m(m^2-1)\delta_{m+n,0} \, .
\nonumber
\end{align}
Classical values of the zero modes $L_0$ and $M_0$ now become eigenvalues 
$h_L =L_0/\hbar$, $h_M=M_0/\hbar$ of the corresponding operators.  The true 
quantum symmetry may be a deformation of (\ref{xxe2})---other central terms may 
appear, for example---but differences will be suppressed by factors of $\hbar$.

Now, it is well known that for a theory with a two-dimensional  conformal 
symmetry, the central charge completely fixes the asymptotic behavior 
of the density of states \cite{Cardy,Cardyb}.  For the simplest case of free bosons and 
fermions, the Cardy formula for the density of states is just the Hardy-Ramanujan formula
for partitions of an integer \cite{Hardy}.  For the general case, I know of no elementary 
explanation; for a careful but not terribly intuitive derivation, see  \cite{Carlipy}.  Roughly 
speaking, exact conformal symmetry is powerful enough to prevent any exponential growth 
in the number of states, which can occur only because of the anomalous symmetry-breaking 
characterized by the central charge $c$.

The BMS${}_3$ symmetry (\ref{xxe2}) is not quite a  conformal symmetry, but as 
Bagchi et al.\ have shown \cite{Bagchi}, it has its own version of the Cardy formula for 
the asymptotic density of states.  In hindsight, this is not so surprising, since BMS${}_3$ can 
be obtained as a contraction of the two-dimensional conformal algebra.  The resulting 
entropy---the logarithm of the density of states at fixed eigenvalues $h_L$ and $h_M$---has
the asymptotic behavior
\begin{align} 
S \sim 
  2\pi h_{\scriptscriptstyle L}\sqrt{\frac{{\hat c}_{\scriptscriptstyle LM}}{2h_{\scriptscriptstyle M}}}
  = \frac{2\pi}{\hbar}L_0\sqrt{\frac{c_{\scriptscriptstyle LM}}{2M_0}} \, ,
\label{xxe3}
\end{align}
where $L_0$, $M_0$, and $c_{\scriptscriptstyle LM}$ in the last equality are the classical 
values.  Note that the factors of Planck's constant combine to give an overall $1/\hbar$,
an expected feature of an entropy described in terms of a  classical phase space. 

\section{Reduction to two dimensions \label{secreduc}}

The first step in our derivation of black hole entropy will be to reduce the problem to two
dimensions.  To understand this process, it is helpful to start with a rather elaborate description 
of an ordinary Schwarzschild black hole in $D$ spacetime dimensions.   The horizon $\Delta$ 
of such a black hole is a Killing horizon, that is, a null $(D-1)$-manifold whose null normal 
coincides with a Killing vector $\chi$.   The integral curves of $\chi$ on 
$\Delta$ are null geodesics, the generators of $\Delta$.   Since $\chi$ is timelike outside the 
horizon, it determines a preferred time coordinate, and through that a foliation of $\Delta$ by
$(D-2)$-spheres $\hat\Delta$ of constant time.

This structure allows us to identify two ``preferred'' directions at any point $p$ on $\Delta$: 
the direction of the  Killing vector at $p$ and the outward radial direction transverse to
$\Delta$ and normal to the slice $\hat\Delta$ containing $p$.   These determine a local 
``$r$--$v$ plane,''  where $r$ is a radial coordinate and $v$ is a parameter along the
null generators of the horizon.  If we further choose coordinates $y^\mu$ on $\hat\Delta$,
then $(v,r,y^\mu)$ can be extended to form a Gaussian null coordinate system near  the horizon  
(see, for instance, Appendix A of \cite{Wald}).  For the Schwarzschild case, these coordinates
are essentially  Eddington-Finkelstein coordinates, with $r$ shifted so that $r=0$ at the horizon.   
By spherical symmetry, each slice $\hat\Delta$ at constant $(v,r)$ is invariant under
rotations, so standard Kaluza-Klein methods can reduce the Einstein-Hilbert action to that of
two-dimensional dilaton gravity \cite{Grumiller}.

The reason I have given such a complicated description of a relatively simple procedure is 
that most of the steps generalize quite broadly.  Let $\Delta$ by a nonexpanding horizon 
\cite{Ashtekar,LewPaw}, that is, a null $(D-1)$-dimensional manifold with null normal
$\ell$ such that
\begin{enumerate}
\item $\Delta$ has the topology $(0,1)\times{\hat\Delta}$, where $\hat\Delta$
is usually taken to be compact (a sphere for a black hole, a torus for a black ring, etc.);
\item The expansion $\theta^{(\ell)}$ of the null normal vanishes.  If the stress-energy
tensor satisfies the null energy condition for $\ell$, that is, $T_{AB}\ell^A\ell^B\ge0$,
then this condition further implies \cite{LewPaw} that $\ell$ has vanishing shear 
and that
\begin{align}
\mathcal{L}_\ell q_{ab}=0 \, ,
\label{xy1}
\end{align}
\end{enumerate}
where $\mathcal{L}$ is the Lie derivative and $q_{ab}$ is the (degenerate) induced
metric on $\Delta$.  

A nonexpanding horizon generalizes the notion of a Killing horizon, dropping the 
requirement of a Killing vector but retaining a time translation
symmetry on the horizon itself.  As in the Schwarzschild case, the integral curves
of $\ell^a$ on $\Delta$ are the null geodesic generators of $\Delta$.  In the absence of
a Killing vector, there is no preferred time coordinate, but it may be shown that a
generic nonexpanding horizon has a preferred foliation, a set of ``good cuts,'' that
generalize the constant time slices of the Schwarzschild metric \cite{Kor}.
Hence we can again construct an ``$r$--$v$ plane'' and a Gaussian null coordinate
system near $\Delta$.

The main change from the Schwarzschild case is that the cross-section $\hat\Delta$ 
of a nonexpanding horizon need not have any symmetries, so conventional 
Kaluza-Klein methods no longer apply.  As Yoon has shown, though, there is a
generalized Kaluza-Klein reduction even in the absence of symmetries \cite{Yoon,Yoonb}.
Start by writing the metric in the general form
\begin{align}
ds^2 = g_{AB}dz^Adz^B =
g_{ab}dx^adx^b + \phi_{\mu\nu}(dy^\mu + A_a{}^\mu dx^a)(dy^\nu + A_b{}^\nu dx^b) \, ,
\label{xy2}
\end{align}
where lower case Roman indices (a,b,\dots) run from $0$ to $1$, lower case Greek
indices ($\mu$,$\nu$,\dots) run from $2$ to $D-1$, and upper case Roman indices ($A$,$B$,\dots)
run from $0$ to $D-1$.  The ``$x$'' coordinates label our 
preferred two-dimensional manifold, while the ``$y$'' coordinates are the transverse directions.  
Define the Kaluza-Klein-like derivatives
\begin{align}
&{\hat\partial}_a = \partial_a - A_a{}^\mu\partial_\mu \, ,\nonumber\\[.3ex]
&D_a\phi_{\mu\nu} = {\hat\partial}_a\phi_{\mu\nu} 
   - (\partial_\mu A_a{}^\rho)\phi_{\rho\nu} -  (\partial_\nu A_a{}^\rho)\phi_{\rho\mu} \, ,
   \label{xy3}
\end{align}
connections
\begin{align}
&{\hat\Gamma}^a_{bc} 
    = \frac{1}{2}g^{ad}\left({\hat\partial}_b g_{dc} +  {\hat\partial}_c g_{db} - {\hat\partial}_d g_{bc}\right) \, ,
    \nonumber\\[.3ex]
&{\hat\Gamma}^\rho_{\mu\nu} 
        = \frac{1}{2}\phi^{\rho\sigma}\left(\partial_\mu \phi_{\sigma\nu} +  \partial_\nu \phi_{\sigma\mu} 
        - \partial_\sigma \phi_{\mu\nu} \right) \, , \label{xy4}
        \end{align}
and curvatures
\begin{align}        
&{\hat R}_{ab} = {\hat\partial}_c{\hat\Gamma}^c_{ab} - {\hat\partial}_a{\hat\Gamma}^c_{bc}
       +{\hat\Gamma}^c_{ab}{\hat\Gamma}^d_{cd} - {\hat\Gamma}^c_{ad}{\hat\Gamma}^d_{bc} \, , \quad
       \nonumber\\[.3ex]
&{\cal R}_{\mu\nu} =  \partial_\rho{\hat\Gamma}^\rho_{\mu\nu} 
       - \partial_\mu{\hat\Gamma}^\rho_{\nu\rho} + {\hat\Gamma}^\rho_{\mu\nu}{\hat\Gamma}^\sigma_{\rho\sigma} 
       - {\hat\Gamma}^\rho_{\mu\sigma}{\hat\Gamma}^\sigma_{\nu\rho} \, , \nonumber\\[.3ex]
&F_{ab}{}^\mu = {\hat\partial}_a A_b{}^\mu - {\hat\partial}_b A_a{}^\mu \, ,  \nonumber\\[.3ex]      
&{\hat R}=  g^{ab}{\hat R}_{ab} \, , \quad  {\cal R}= \phi^{\mu\nu} {\cal R}_{\mu\nu} \, .\label{xy5}
\end{align}
A straightforward calculation of the Einstein-Hilbert action then gives
\begin{align}
I =  \int d^{D-2}y\,I_2
\label{xy6}
\end{align}
with
\begin{align}
I_2 = \frac{1}{16\pi G} \int d^2x\! \sqrt{-g}\sqrt{\phi} \Biggl\{ 
      {\hat R} &+ \frac{1}{4}g^{ab}g^{cd}\phi_{\mu\nu}F_{ac}{}^\mu F_{bd}{}^\nu   \nonumber\\[.3ex]
      &+ \frac{1}{4}g^{ab}\phi^{\mu\nu}\phi^{\rho\sigma}\left( D_a\phi_{\mu\rho}D_b\phi_{\nu\sigma}
      - D_a\phi_{\mu\nu}D_b\phi_{\rho\sigma} \right) \nonumber\\[.3ex]
      &+ \frac{1}{4}\phi^{\mu\nu}g^{ab}g^{cd}\left( \partial_\mu g_{ac}\partial_\nu g_{bd} 
       - \partial_\mu g_{ab}\partial_\nu g_{cd} \right) + {\cal R}  \Biggr\} \, ,
       \label{xy7}
\end{align}
where $g$ and $\phi$ are the determinants of $g_{ab}$ and $\phi_{\mu\nu}$.

The first two lines in (\ref{xy7}) look like an ordinary Kaluza-Klein reduction, and can in fact
be viewed as the action of a Kaluza-Klein theory whose gauge group is the group of 
diffeomorphisms of the transverse manifold \cite{Yoon}.  A two-dimensional interpretation  
of the third line is less obvious.  We can cure this, though, with a partial gauge fixing.  First, 
we can always choose local coordinates in which
\begin{align}
g_{ab} = \left(\begin{array}{rr} -2h & 1 \\ 1 & 0 \end{array}\right) \, .
\label{xy8}
\end{align}
Yoon calls this ``Polyakov gauge,'' after a similar choice in two-dimensional field
theories \cite{Polyakov}, while from the $D$-dimensional point of view it is essentially
Bondi gauge \cite{Bondi} or Gaussian null coordinates \cite{Wald}.  It is easy to see
that with this choice, even if $h$ depends on the $y^\mu$,
\begin{align}
\phi^{\mu\nu}g^{ab}g^{cd}\left( \partial_\mu g_{ac}\partial_\nu g_{bd} 
       - \partial_\mu g_{ab}\partial_\nu g_{cd} \right) = 0 \, .
\label{xy9}
\end{align}
For our purposes, (\ref{xy8}) is too restrictive a gauge choice---it hides a piece of the symmetry 
we are trying to understand.  But if we now allow an arbitrary two-dimensional coordinate 
transformation $x\rightarrow{\bar x}(x)$, it may be checked that the term (\ref{xy9}) still vanishes.

The remaining term of concern in (\ref{xy7}) is the transverse curvature $\cal R$.  
In $D=4$ dimensions, this term is essentially trivial:
\begin{align}
\int d^2y \sqrt{\phi}{\cal R} = 4\pi\chi \, ,
\label{xy10}
\end{align}
where $\chi$ is the Euler characteristic of the transverse manifold.  The transverse
curvature thus merely contributes an effective cosmological constant to the
two-dimensional action.  For $D\ne4$, the situation is more complicated; the transverse 
curvature couples only to $\sqrt{-g}$, but it can give a sort of position-dependent cosmological 
``constant.''   This should not affect the symmetries derived in section \ref{secsym}, but a deeper
understanding would be helpful.   (Note also that $\cal R$ involves only $y$ derivatives, 
while we have chosen a gauge in which $\sqrt{-g}$ depends only on $x$, so by rescaling
$\phi_{\mu\nu}$ by an appropriate power of $\sqrt{-g}$ we can actually remove the
coupling.)

To make it easier to compare this formalism to other work on dilaton gravity, it is
 convenient to separate out the determinant $\phi$ from the transverse metric
$\phi_{\mu\nu}$, writing  
\begin{align}
\phi_{\mu\nu} = \phi^{\frac{1}{D-2}}\Phi_{\mu\nu} \, , \quad \det\left|\Phi_{\mu\nu}\right|=1 \, .
\label{xy11}
\end{align}
The determinant and $\Phi_{\mu\nu}$ then decouple in the kinetic term in (\ref{xy7}),
\begin{align}
\frac{1}{4}g^{ab}&\phi^{\mu\nu}\phi^{\rho\sigma}\left( D_a\phi_{\mu\rho}D_b\phi_{\nu\sigma}
      - D_a\phi_{\mu\nu}D_b\phi_{\rho\sigma} \right) \nonumber\\[.3ex]
     & = \frac{1}{4}g^{ab}(\Phi^{-1})^{\mu\nu}(\Phi^{-1})^{\rho\sigma}\left( D_a\Phi_{\mu\rho}D_b\Phi_{\nu\sigma}
      - D_a\Phi_{\mu\nu}D_b\Phi_{\rho\sigma} \right)
      - \frac{D-3}{4(D-2)}\phi^{-2}g^{ab}D_a\phi D_b\phi \, .
\label{xy12}
\end{align}
One more simplification is standard in dilaton gravity: by rescaling the metric $g_{ab}$, we 
can eliminate the kinetic term for $\phi$.  Specifically, if we set 
\begin{align}
\phi = \varphi^2
\label{xyz1}
\end{align}
and let
\begin{align}
g_{ab} = \varphi^{-\frac{D-3}{D-2}}\,{\bar g}_{ab} \, ,
\label{xy13}
\end{align}
the action (\ref{xy7}) reduces to
\begin{align}
I_2 = \frac{1}{16\pi G} \int d^2x &\sqrt{-{\bar g}}\, \Biggl\{
       \varphi {\bar R} + \frac{1}{4}\varphi^2\,{\bar g}^{ab}{\bar g}^{cd}\Phi_{\mu\nu}
      F_{ac}{}^\mu F_{bd}{}^\nu \nonumber\\[.3ex]
      &\ \ + \frac{1}{4}\varphi\, {\bar g}^{ab}(\Phi^{-1})^{\mu\nu}(\Phi^{-1})^{\rho\sigma}
      \left( D_a\Phi_{\mu\rho}D_b\Phi_{\nu\sigma}
      - D_a\Phi_{\mu\nu}D_b\Phi_{\rho\sigma} \right) + \varphi^{\frac{1}{D-2}}{\cal R}  \Biggr\} \, ,
        \label{xy14}
\end{align}
which may be recognized as the action for a gauge field and a nonlinear sigma model coupled 
to two-dimensional dilaton gravity.

I have, of course, glossed over an essential feature: the ``two-dimensional'' fields in (\ref{xy14})
also depend on the transverse coordinates $y$, which must still be integrated over.  This
remnant of the higher dimensional structure appears in two places.  First, the ``two-dimensional'' 
curvature $R_{ab}$ in (\ref{xy5}) involves convective derivatives ${\hat\partial}_a 
= \partial_a - A_a{}^\mu\partial_\mu$, distinguishing it from the ordinary Ricci curvature of the 
two-dimensional metric ${\bar g}_{ab}$.  Near a black hole horizon $\Delta$,  though, one can 
choose corotating coordinates in which $A_a{}^\mu$ vanishes on the horizon and  remains 
small in a neighborhood of $\Delta$ \cite{Wald}.   We will be interested in symmetries in a 
small region around the horizon.  In such a region, to the order of approximation we will
need, these coordinates will allow us to replace $\bar R$ in (\ref{xy14}) by the ordinary 
two-dimensional curvature scalar (see \ref{AppB} for details).

Second, the ``matter'' fields $F_{ab}{}^\mu$ and $\Phi_{\mu\nu}$ in (\ref{xy14}) also depend
on the transverse coordinates.  In a symmetric enough setting, we could expand these fields 
in modes to create a Kaluza-Klein tower of states.  In general, though---for instance, for a black 
hole whose horizon is distorted by surrounding matter---this will not be possible.  Fortunately, 
though, it is also not necessary.  As Wall showed for quantum fields on a black hole horizon 
\cite{Wall}, the physics at different transverse positions decouples, and each null generator 
can be treated separately.  While I do not know a rigorous generalization to the case of
dynamical gravity, we shall see that the relevant symmetries act separately on each
generator.  Since these symmetries govern the density of states, this density is also
determined independently on each generator, and the total entropy can be obtained
by integrating.  This is the underlying reason for an area law for entropy, although as we
shall see, it gives more, fixing the exact coefficient of the area.

Note that we have not yet imposed the existence of a horizon.  Even the use of Gaussian null
coordinates requires only the presence of a null surface, which need not have vanishing
expansion.  In principle, it is possible to further restrict the metric (\ref{xy2}), but the resulting
expressions are complicated and unwieldy \cite{Booth}.  We will instead take a shortcut,
identifying horizons in the two-dimensional action $I_2$ to obtain a more tractable formulation.

\section{Dilaton gravity with null dyads \label{secdil}}

We now restrict our attention to the effective two-dimensional action (\ref{xy14}) at fixed
transverse position (fixed $y$).  While this action will not reveal the full symmetries of the 
higher-dimensional theory, any $y$-independent symmetry of $I_2$ will also be a
symmetry of the full theory.

To simplify notation, let us write
\begin{align}
I = \frac{1}{16\pi G}\int_M\!\left(\varphi R + V[\varphi,\chi] \right){\epsilon} \, ,
\label{a01}
\end{align}
where $\epsilon$ is the volume two-form\footnote{I am using the convention that the object 
one integrates over an $n$-manifold is an $n$-form \cite{Waldtext}.  For our two-dimensional 
manifold, the volume form is $\epsilon_{ab}$.  For a null line with null normal $\ell_a$, the volume 
one-form is $n_a$, a null vector normalized so that $\ell\cdot n=-1$.}  and $\chi$ denotes any further 
fields in the problem (here $\Phi$, $A$, and any additional matter fields).  The quantity $\varphi$ 
is called the dilaton; as we have seen, it is essentially the volume element of the 
transverse metric.  Note that while the potential $V$ in (\ref{a01}) may be quite complicated, it 
contains no $x$ derivatives of $\varphi$.

The equations of motion coming from varying $g$ and $\varphi$ in this action are
\begin{subequations}
\begin{align}
&E_{ab} = \nabla_a\nabla_b\varphi - g_{ab}\Box\varphi + \frac{1}{2}g_{ab}V = 8\pi G T_{ab} \, ,
   \label{ab5}\\
&R + \frac{dV}{d\varphi} = 0 \, , \label{ab5b}
\end{align}
\end{subequations}
where I have added a source stress-energy tensor.  Equation (\ref{ab5b}) is not independent, but 
follows from the divergence of (\ref{ab5}).  

It is convenient to describe the geometry in terms of a null dyad $(\ell_a,n_a)$, with 
$\ell^2= n^2 = 0$ and $\ell\cdot n=-1$.    In terms of such a dyad, the metric and volume form are
\begin{align}
g_{ab} = - \left(\ell_an_b + n_a\ell_b\right) \, , \quad 
\epsilon_{ab} =  \left(\ell_an_b - n_a\ell_b\right) \, .
\label{a1}
\end{align}
When we later specialize to the case of a black hole spacetime with horizon $\Delta$, we will 
choose a dyad for which $\ell$ is the null normal to $\Delta$ and $n$ is the induced volume
element.  

To simplify later equations, we define derivatives
\begin{align}
D=\ell^a \nabla_a \, , \quad {\bar D} = n^a \nabla_a \, .
\label{a1a}
\end{align}
$D$ is essentially the same $D$ as in the Newman-Penrose formalism.  $\bar D$ would 
ordinarily be denoted $\Delta$ in the Newman-Penrose formalism, but we are already using 
$\Delta$ to signify the horizon.

The dyad $(\ell, n)$ is determined only up to local Lorentz transformations, 
\begin{align}
\ell^a \rightarrow e^\lambda\ell^a, \quad n^a  \rightarrow e^{-\lambda} n^a \, .
\label{a1b}
\end{align}
We can partially fix this freedom by choosing $n_a$ to have vanishing acceleration,
$n^b\nabla_bn^a = 0$.  This condition implies that the integral curves of $n$ are 
affinely parametrized null geodesics, which can be taken to start at the horizon $\Delta$.
The symmetry (\ref{a1b}) is still not completely fixed, but  the remaining transformations are 
restricted to those for which ${\bar D}\lambda=0$.  

With this condition on $n_a$,  it is easy to check that
\begin{alignat}{3}
&\nabla_a\ell_b = - \kappa n_a\ell_b \, , \qquad\qquad
   && \nabla_a\ell^a = \kappa \, , \nonumber\\
&\nabla_a n_b =  \kappa n_a n_b \, , && \nabla_a n^a =0 \, ,
\label{a2}
\end{alignat}
where $\kappa$ will later be interpreted as the surface gravity at a horizon.  By (\ref{a1}),
$\ell_a$ is a conformal Killing vector.  Under variation of the dyad,  (\ref{a2}) will be preserved
as long as
\begin{align}
&{\bar D}(\ell^c\delta n_c) = (D+\kappa)(n^c\delta n_c) \, ,\nonumber\\
&\delta\kappa = -  D(n^c\delta\ell_c) +  \kappa \ell^c\delta n_c 
   +  {\bar D}(\ell^c\delta\ell_c) \, .
\label{a3a}
\end{align}

We will later need to integrate by parts along the horizon.  For this, it will  be useful to
take advantage of the identity
\begin{align}
(df)_a = -Df\,n_a - {\bar D}f\,\ell_a \quad\hbox{for any function $f$,}  \label{abc1}
\end{align}
where I am treating $n_a$ and $\ell_a$ as one-forms.  Another identity will also be helpful:
\begin{align}
[D,{\bar D}] = -\kappa{\bar D}\quad  \Leftrightarrow\quad {\bar D}D=(D+\kappa){\bar D} \, .
\label{ab4} 
\end{align}

By considering the commutator $[\nabla_a,\nabla_b]\ell^b$ and recalling that in
two dimensions $R_{ab} = \frac{1}{2}g_{ab}R$, it is straightforward to show
from (\ref{a2}) that
\begin{align}
R =  2{\bar D}\kappa \, .
\label{a4}
\end{align}
The action (\ref{a01}) can thus be written as
\begin{align}
I = \frac{1}{8\pi G}\int_M\!\left(-\kappa{\bar D}\varphi + \frac{1}{2}V[\varphi,\chi] \right){\epsilon} \, .
\label{a01b}
\end{align}

\section{Horizons \label{sechor}}

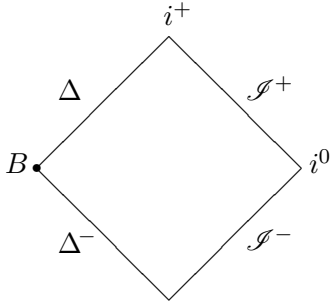
\begin{figure}
\centering
\begin{picture}(100,100)(0,-45)
\put(0,0){\line(1,1){50}}
\put(0,0){\line(1,-1){50}}
\put(100,0){\line(-1,1){50}}
\put(100,0){\line(-1,-1){50}}
\put(8,26){$\Delta$}
\put(79,26){$\mathscr{I}^+$}
\put(48,54){$i^+$}
\put(8,-32){$\Delta^{\!-}$}
\put(79,-31){$\mathscr{I}^-$}
\put(103,-2){$i^0$}
\put(-12,-2){$B$}
\put(0,0){\circle*{3}}
\end{picture}
\caption{Typical Penrose diagram for the exterior of a black hole \label{fig1}}
\end{figure}

Our next task will be to characterize a generic black hole horizon in this two-dimensional 
setting, as a first step toward analyzing its near-horizon symmetries. Spacetimes 
containing black holes in two-dimensional dilaton gravity have essentially
the same Penrose diagrams as those in higher dimensions \cite{Grumiller}, 
as illustrated in figure \ref{fig1}.    As in higher dimensions, the horizons are Killing horizons
\cite{Kunstatter}, and form boundaries of trapped regions \cite{Cai}.  For simplicity,
figure \ref{fig1} shows an asymptotically flat black hole.  This asymptotic behavior will be
irrelevant for the main argument of this paper, though; the analysis below will hold equally 
well for asymptotically de Sitter or anti-de Sitter black holes.

Let us choose our null dyad so that at the horizon, $\ell$ coincides with the null
normal to the horizon.  This means $\ell$ is also tangent to the horizon---its
inner product with the normal (itself) is zero.  Indeed, in any dimension the null
normals to the horizon are the tangent vectors of the null generators of the
horizon.  (See the beginning of \cite{Gour} for a nice review.)
 
In two dimensions, this choice of dyad is straightforward.  The lift to $D$ dimensions, 
though, is  ambiguous; depending on coordinate choices, $\ell_A$ and $\ell^A$ 
may have additional transverse  components.  Our philosophy here will be that our
preferred two-dimensional subspace traces the generators of the horizon, that is,
that each generator of the horizon $\Delta$ has constant transverse coordinates $y$.
This means that $\ell$ \emph{as a tangent vector} lies in our two-dimensional subspace; 
that is, $\ell^\mu=0$ but $\ell_\mu$ need not vanish.  We will choose $n$ so that 
$n_\mu=0$.  This ensures that even from the higher dimensional point of view, $n_a$ 
is the volume one-form along a horizon generator.  As long as we stick to the coordinates
described at the end of section \ref{secreduc}, in which $g_{\mu a}=0$ at the horizon,
these choices will be largely irrelevant, but they may be important in more general
coordinate systems. 

We now need a way to determine that $\Delta$ is indeed a horizon.  This will require 
an appropriate generalization of the ``non-expanding horizon'' criteria of section 
\ref{secreduc}.   From the $D$-dimensional point of view, the expansion
of $\Delta$ is
\begin{align}
\theta = \left(\delta^A_B + \ell^A n_B \right)\nabla_A\ell^B
  = \left(\delta^a_b + \ell^a n_b\right)\nabla_a\ell^b + \nabla_\mu\ell^\mu 
  = \ell^a\Gamma^\mu_{a\mu} \, .
\label{c1xx}
\end{align}
In general, this will be a complicated expression, involving both the transverse metric
$\phi_{\mu\nu}$ and the mixed components $A_a{}^\mu$.  But recall that we have
chosen coordinates in which $A_a{}^\mu \tq 0$, where from now on I will use the symbol $\tq$ 
to mean ``equal on the horizon.''  Hence
\begin{align}
\theta = \ell^a\Gamma^\mu_{a\mu} \tq \frac{1}{2}\phi^{\mu\nu}\ell^a{\partial}_a\phi_{\mu\nu}
  = \frac{1}{\varphi}D\varphi \, .
  \label{c1x}
\end{align}
The condition for vanishing expansion is thus $D\varphi=0$, and we will use this as the
means to locate the horizon.  Near the horizon, $D\varphi$ can then serve as a small 
expansion parameter, indicating how far we have moved from $\Delta$.  The interpretation
of this parametrization in terms of Gaussian null coordinates is described in \ref{AppB}.

In higher dimensions, vanishing expansion is enough to ensure that the whole horizon
geometry is stationary as well.  In two dimensions, where the dilaton $\varphi$
is now separate from the transverse metric, this is no longer the case, and we must separately
require that $DR \tq 0$.  I will also impose one more boundary condition at the horizon, that  
the integration measure $n_a$ remain fixed at $\Delta$.   This restriction appears to be needed 
for the covariant canonical symplectic form of section \ref{secsympstruc} to be well-behaved,
though further exploration would be interesting.  In \cite{Carlipxa}, the condition $\ell^a\delta\ell_a=0$ 
was also imposed, but while this simplifies the symplectic structure, it is not really needed.

Our boundary conditions at $\Delta$ thus become
\begin{subequations}
\begin{align}
&D\varphi \tq 0 \, , \label{cxa}\\[.3ex]
&DR \tq 0 \, , \label{cxb}\\[.3ex]
&\ell^a\delta n_a \tq n^a\delta n_a \tq 0 \label{cxc}
\end{align}
\end{subequations}
(where, again, $\tq$ means``equal on $\Delta$'').
Our task is to find the symmetries of the dilaton gravity action that are compatible with 
these conditions.  

Before proceeding further, one slightly subtle issue of interpretation 
should be addressed.  The approach here is not to first choose a
fixed submanifold $\Delta$ and then impose (\ref{cxa})--(\ref{cxc}).  This is too strong
a demand: it would forbid variations that changed $D\varphi$ on this fixed surface,
and would prohibit transverse diffeomorphisms at $\Delta$.  The philosophy is,
rather, to use the condition $D\varphi=0$ to determine the location of the horizon,
and to impose the remaining conditions at that location.   A variation that changes $D\varphi$ 
is then understood as changing the location of the horizon, and a suitable transverse 
diffeomorphism can be used to ``move it back.''

\section{Horizon symmetries \label{secsym}}

The action (\ref{a01}) is, of course, invariant under two-dimensional diffeomorphisms, 
including horizon ``supertranslations'' \cite{Donnay} generated by vector fields 
$\xi^a = \xi\ell^a$.   Such diffeomorphisms fail to respect condition (\ref{cxc}), however,
since $\ell^a\delta_\xi n_a \ne 0$.  This is easily cured, by supplementing each
diffeomorphism with a local Lorentz transformation $\delta\ell^a = (\delta\lambda)\ell^a$,
$\delta n^a = -(\delta\lambda)n^a$ with $\delta\lambda = D\xi$.  From the discussion after
(\ref{a1b}), this requires that ${\bar D}\xi\tq0$.   We thus have an invariance
\begin{align}
&\delta_\xi\ell^a = 0 \, , \quad \delta_\xi n^a = -(D+\kappa)\xi \,n^a \, ,\nonumber\\
&\delta_\xi g_{ab} = (D+\kappa)\xi \, g_{ab} \, ,\nonumber\\
&\delta_\xi \varphi = \xi D\varphi \, , \quad \hbox{with ${\bar D}\xi\tq0$} \, .
\label{c1}
\end{align}
Note that from (\ref{a2}),
\begin{align}
\delta_\xi\kappa = D(D+\kappa)\xi \, .
\label{c1xy}
\end{align}

As pointed out some time ago \cite{Carlip6,Kang}, for configurations containing black holes
the action also has an \emph{approximate} invariance under certain shifts of the 
dilaton near the horizon, with an approximation that can be made arbitrarily good 
by restricting the transformation to a small enough neighborhood of $\Delta$. 
This is not quite an ordinary invariance, since it holds only for a restricted class of 
configurations, those with horizons.   For such configurations, though, it can be made
arbitrarily close to an exact symmetry (see \ref{AppC} for more details).

Specifically, consider a variation 
\begin{align}
{\hat\delta}_\eta\varphi = \nabla_a(\eta\ell^a) 
   = (D+\kappa)\eta  \quad \hbox{with ${\bar D}\eta\tq0$} \, .
\label{c1a}
\end{align}
(The hat on $\hat\delta$ distinguishes this variation from a diffeomorphism.)
The action transforms as 
\begin{align}
{\hat\delta} _\eta I &= \frac{1}{16\pi G}\int_M\!  \left(R + \frac{dV}{d\varphi}\right)
  {\hat\delta}_\eta\varphi\, {\epsilon}  
 = -\frac{1}{16\pi G}\int_M \eta\left[ DR + \frac{d^2V}{d\varphi^2}D\varphi  
   \right]{\epsilon} \, .
\label{c2}
\end{align}
But $D\varphi$ and $DR$ both vanish at the horizon, so  the variation (\ref{c2}) 
can be made as small as one wishes by choosing $\eta$ to fall off fast enough away 
from $\Delta$.   

There is one subtlety, however.  While the transformation (\ref{c1a}) does not directly 
act on the curvature, the change of $\varphi$ ``moves the horizon''---that is, the locus 
$D\varphi=0$ may change under a shift of $\varphi$.  In itself, this is not a problem, but  
$DR$ may no longer vanish at the new location.  The diffeomorphism needed to 
``move the horizon back'' is determined by the condition  
\begin{align}
({\hat\delta}_\eta + \delta_\zeta)(D\varphi) =
{\hat\delta}_\eta(D\varphi) + \zeta^a\nabla_a(D\varphi) \tq 0 \ \Rightarrow \ 
\zeta^a = {\bar\zeta}n^a = -\frac{D({\hat\delta}_\eta\varphi)}{{\bar D}D\varphi}n^a \, .
\label{ac8}
\end{align}
This change can be compensated  with a ``small''  (order $D\varphi$) Weyl transformation
 of the metric to restore the condition $DR\tq0$.  Consider a transformation of the form
\begin{align}
{\hat\delta}_\eta g_{ab} = {\hat\delta}\omega_\eta\,g_{ab} \quad \Leftrightarrow\quad
       {\hat\delta}_\eta \ell_a = {\hat\delta}\omega_\eta\,\ell_a \, ,
\label{c7z}
\end{align}
where the second equality comes from the boundary condition that $\delta n_a\tq0$.       
Define
\begin{align}   
{\hat\delta}\omega_\eta = X_\eta\frac{D\varphi}{{\bar D}D\varphi} \, .
\label{c7}
\end{align}
Using identities from section \ref{secdil}, it is not hard to see that $R$ transforms as
\begin{align}
{\hat\delta}_\eta R  \tq 2(D+\kappa)X_\eta \, .
\label{c7z1}
\end{align}
The condition that $DR$ remain zero on $\Delta$ is thus
\begin{align}
{\bar\zeta}{\bar D}DR + 2D(D+\kappa)X_\eta \tq 0 \, .
\label{c7zz}
\end{align} 
On shell---or, less restrictively, whenever the constraint (\ref{ab5b}) holds---a short calculation 
gives an explicit expression for $X_\eta$:
\begin{align}
{\bar\zeta}{\bar D}DR + 2D(D+\kappa)X_\eta &\tq
  \frac{d^2V}{d\varphi^2}D(D+\kappa)\eta + 2D(D+\kappa)X_\eta \tq 0 \ \Rightarrow \
   X_\eta \tq -\frac{1}{2}\frac{d^2V}{d\varphi^2}\eta \, .
\label{cx2}
\end{align}
With this added transformation, the boundary condition $DR\tq0$ is preserved.
Like (\ref{c1a}), the Weyl transformation (\ref{c7})  changes the action only by terms 
proportional to $\eta D\varphi$, which can be made arbitrarily small by choosing $\eta$
to fall off fast enough away from the horizon.

We thus have two sets of transformations at the horizon, diffeomorphisms $\delta_\xi$
and shifts ${\hat\delta}_\eta$, which preserve the action (to an arbitrarily good 
approximation) as long as a horizon actually exists.  It is not too hard to check
that these satisfy an algebra
\begin{alignat}{3}
&[\delta_{\xi_1}, \delta_{\xi_2}] f \tq \delta_{\xi_{12}}f \qquad
  && \hbox{with}\ \ \xi_{12} = (\xi_1D\xi_2 - \xi_2D\xi_1) \, , \nonumber\\
&[{\hat\delta}_{\eta_1}, {\hat\delta}_{\eta_2}] f \tq 0 \, , \label{d1}\\
&[\delta_{\xi_1}, {\hat\delta}_{\eta_2}] f \tq {\hat\delta}_{\eta_{12}}f \qquad
  && \hbox{with}\ \ \eta_{12} = - (\xi_1D\eta_2 - \eta_2D\xi_1) \, . \nonumber
\end{alignat}
This may be recognized as a BMS$_3$ algebra, or equivalently a Galilean conformal algebra 
\cite{GCA}.

Given the rather atypical nature of this shift symmetry, 
we should also check the variation of the equations of motion (\ref{ab5})--(\ref{ab5b}).
These are, of course, preserved by diffeomorphisms, so we need only consider the
transformations (\ref{c1a}) and (\ref{c7}).  Since we are assuming that $\eta$ falls off rapidly 
away from the horizon, it is enough to check the variations at $\Delta$.  
By a straightforward computation, most of the equations of motion are preserved: up
to terms that are themselves proportional to the equations of motion,
\begin{subequations}
\begin{align}
&g^{ab}{\hat\delta}_\eta E_{ab} \tq 2(D+\kappa){\bar D}{\hat\delta}_\eta\varphi 
   +  \frac{dV}{d\varphi} {\hat\delta}_\eta\varphi 
    \tq \left(R + \frac{dV}{d\varphi}\right)(D+\kappa)\eta \, , \label{ce2}\\
&n^an^b{\hat\delta}_\eta E_{ab} \tq {\bar D}^2{\hat\delta}_\eta\varphi
  - {\bar D}\varphi{\bar D}{\hat\delta}_\eta\omega 
  \tq \frac{1}{2}{\bar D}\left(R + \frac{dV}{d\varphi}\right)\eta \, , \label{ce1}\\
&{\hat\delta}_\eta\left( R + \frac{dV}{d\varphi}\right) 
   \tq {\hat\delta}_\eta R + \frac{d^2V}{d\varphi^2}{\hat\delta}_\eta\varphi \, ,
   \label{ce4}
\end{align}
\end{subequations}
where I have used (\ref{ab4}), (\ref{a4}), and the condition ${\bar D}\eta\tq 0$.

The remaining variation, $\ell^a\ell^b{\hat\delta}_\eta E_{ab}$, is not zero.  But this is 
actually a familiar occurrence in conformal field theory.  If we set $E_{ab} = 8\pi G T_{ab}$,  
we find that
\begin{align}
\ell^a\ell^b{\hat\delta}_\eta T_{ab} \tq \frac{1}{8\pi G} (D-\kappa)D(D+\kappa)\eta \, ,
\label{ca6}
\end{align}
which is essentially the usual anomaly for a conformal field theory with a central charge 
proportional to $1/G$ \cite{CFT}.  This is our first hint that the symmetry is 
anomalous.  

One might worry that this anomaly could spoil the covariant phase 
space construction of \ref{secapp}, since the closure of the symplectic current (\ref{ab6}) 
relies on the classical field equations.  Fortunately, this is not a problem: the only dangerous 
term in the exterior derivative (\ref{ab7a}) is proportional to $n_a n_b\delta g^{ab}$, which  
vanishes on $\Delta$ by virtue of the boundary conditions (\ref{cxc}).

\section{Symplectic structure and generators \label{secsympstruc}}

To complete the analysis of the symmetries of section \ref{secsym}, we should ask whether 
the algebra (\ref{d1}) can be realized---perhaps with a central extension---as a Poisson 
algebra of canonical generators of the symmetries, since this is the formulation that translates 
most directly into quantum mechanics.   We have so far avoided introducing explicit 
coordinates.  We will continue to do so, by employing the covariant canonical formalism 
reviewed in \ref{secapp}.

The symplectic form (\ref{ab7}) is defined as an integral over a Cauchy surface $\Sigma$.
To study horizon symmetries in the covariant phase space formalism, we should incorporate
$\Delta$ as part of our Cauchy surface.  Let us focus on the exterior region of an asymptotically
flat black hole, with a Penrose diagram given by figure \ref{fig1}, and take $\Sigma$ to be the union 
of the future horizon $\Delta$ and future null infinity $\mathscr{I}^+$, with ends at the 
bifurcation point $B$ and spacelike infinity.   As noted earlier, the details of $\mathscr{I}^+$ 
will be unimportant, since we will be considering transformations that are nonvanishing only
in a small neighborhood of the horizon.   

Applying the general relations (\ref{ab6})--(\ref{ab7}) to the action (\ref{a01b}) for dilaton
gravity and using the boundary condition $\delta n_a\tq0$, it is straightforward 
to show that\footnote{The calculation simplifies if one notes that in differential form notation,
$\kappa{\bar D}\varphi\,\epsilon = {\bar D}\varphi\,d\ell = -\kappa\, d\varphi\wedge n$.}
\begin{align}
\Omega_\Delta[(\varphi,g);\delta_1(\varphi,g),\delta_2(\varphi,g)] 
    = \frac{1}{8\pi G}\int_\Delta \left[ \delta_1\varphi\,\delta_2\kappa 
     - \delta_1({\bar D}\varphi)\ell^b\delta_2\ell_b \right] n_a - (1\leftrightarrow 2) \, .
\label{ac7}
\end{align}
The full symplectic form will include an additional integral along $\mathscr{I}^+$, but
this will be irrelevant to our consideration of near-horizon symmetries.

Two  slightly tricky points remain, though, both related to the fact that a variation of 
$\varphi$ can ``move the horizon,'' changing the locus of points $D\varphi=0$.
First, as discussed in \ref{secapp}, the symplectic form itself is independent of the 
integration contour as long as the endpoints remain fixed.  But $\Omega_\Delta$ 
\emph{can} change under variations that move the ends of the Cauchy surface.  
To avoid this behavior, we will require that $\delta(D\varphi)=0$ at the bifurcation point 
$B$ of figure \ref{fig1}, a condition that will be used in section \ref{secmodes}.
  
Second, while typical changes in the horizon locus will not affect $\Omega_\Delta$,  they
will change objects such as Hamiltonians defined as integrals over $\Delta$.  We will 
account for this effect by adding a transverse diffeomorphism to ``move the 
horizon back.''  As in section \ref{secsym}, such a diffeomorphism is determined by the 
condition that
\begin{align}
(\delta + \delta_\zeta)(D\varphi) =
\delta(D\varphi) + \zeta^a\nabla_a(D\varphi) \tq 0 \ \Rightarrow \ 
\zeta^a = {\bar\zeta}n^a 
   = -\left(\frac{D\delta\varphi}{{\bar D}D\varphi} 
   + \frac{{\bar D}\varphi}{{\bar D}D\varphi}\ell^b\delta\ell_b\right)n^a \, .
\label{ac8z}
\end{align}
Hence for an object of the form $H = \int_\Delta \mathscr{H}\,n_a$, the full variation will be
\begin{align}
\delta \int_\Delta \mathscr{H}\,n_a 
   = \int_\Delta( \delta\mathscr{H} + \zeta^a\nabla_a \mathscr{H})n_a \, .
\label{ac9}
\end{align}

We can now ask whether the transformations $\delta_\xi$ and ${\hat\delta}_\eta$ of
the preceding section can be realized canonically as in (\ref{ab8}), that is, whether 
there exist generators that satisfy
\begin{subequations}
\begin{align}
\delta L[\xi] &= \frac{1}{8\pi G}\int_\Delta\left[ \delta\varphi\,\delta_\xi\kappa 
    - \delta_\xi\varphi\,\delta\kappa - \delta({\bar D}\varphi)\ell^b\delta_\xi\ell_b
    + \delta_\xi({\bar D}\varphi)\ell^b\delta\ell_b \right] n_a \nonumber\\
    &= \frac{1}{8\pi G}\int_\Delta\left[ \delta\varphi\,D(D+\kappa)\xi 
    - \xi D\varphi\,\delta\kappa +\left\{ \xi{\bar D}D\varphi 
    - (D+\kappa)\xi{\bar D}\varphi\right\}\!\ell^b\delta\ell_b\right] n_a  \, , \label{d2} \\
\delta M[\eta] &= \frac{1}{8\pi G}\int_\Delta\left[ \delta\varphi\,{\hat\delta}_\eta\kappa 
    - {\hat\delta_\eta}\varphi\,\delta\kappa - \delta({\bar D}\varphi)\ell^b{\hat\delta}_\eta\ell_b
    + {\hat\delta}_\eta({\bar D}\varphi)\,\ell^b\delta\ell_b \right] n_a \nonumber\\
    &= \frac{1}{8\pi G}\int_\Delta\left[ 
    - \delta\omega_\eta\, D\delta\varphi -\delta\kappa(D+\kappa)\eta 
    + \left\{ {\bar D}(D+\kappa)\eta -\delta\omega_\eta\,{\bar D}\varphi\right\}\!\ell^b\delta\ell_b
    \right]n_a \, ,
\label{d6}
\end{align}
\end{subequations}
where in the last line I have used the fact that ${\hat\delta}_\eta\kappa = D{\hat\delta}\omega_\eta$ .

It is not at all clear that such generators exist: there is no obvious reason that the near-horizon
symmetry (\ref{c1a}) should have a canonical realization.  In fact, though, the quantities
\begin{subequations}
\begin{align}
L[\xi] &= \frac{1}{8\pi G}\int_\Delta\left[\xi D^2\varphi - \kappa\xi D\varphi\right]n_a \, ,
\label{d3} \\
M[\eta] &= \frac{1}{8\pi G}\int_\Delta \eta
\left(D\kappa - \frac{1}{2}\kappa^2\right)n_a
\label{d7}
\end{align}
\end{subequations}
do the job.  (To obtain the $\delta\omega_\eta$ terms in (\ref{d6}), one must use the full variation 
(\ref{ac9}), along with equation (\ref{cx2}) for $X_\eta$ and the fact that ${\bar D}\eta\tq 0$; again, 
the covariant phase space formalism allows us to impose equations of motion after variation.)

Using (\ref{ab10}), we can now find the Poisson brackets of these generators:
\begin{subequations}
\begin{align}
&\left\{L[\xi_1],L[\xi_2]\right\} = L[\xi_{12}] \, , \label{d11a}\\
&\left\{M[\eta_1],M[\eta_2]\right\} \tq 0 \, , \label{d11b}\\
&\left\{L[\xi_1],M[\eta_2]\right\} \tq -M[\eta_{12}] 
   - \frac{1}{16\pi G}\int_\Delta \left(D\xi_1 D^2\eta_2 - D\eta_2 D^2\xi_1\right)n_a \, ,
\label{d11c}
\end{align}
\end{subequations}
where $\xi_{12}$ and $\eta_{12}$ were defined in (\ref{d1}).    The $\{L,L\}$ bracket are 
unchanged even if $\Delta$ is not a horizon.  The $\{L,M\}$ and $\{M,M\}$ brackets do
change---the shift transformations are exact symmetries only on a horizon---but modulo equations 
of motion, the deviations are of order $(D\varphi)^2$.  The canonical generators thus give a 
representation of the symmetry algebra (\ref{d1}), now with an added central term.  Such 
central terms are well-understood in classical mechanics \cite{Arnold}; their appearance in 
quantum gravity was first emphasized by Brown and Henneaux \cite{Brown}, and as we saw 
in section \ref{secBMS}, they play a crucial role in determining entropy.   

\section{Modes and zero-modes \label{secmodes}}

As described in section \ref{secBMS}, we can now use the symmetry (\ref{d11a})-- (\ref{d11c}) 
to determine the density of states.  To do so, we will need the central charge and the zero 
modes.  These, in turn, require a mode expansion for the parameters $\xi$ and $\eta$.

For a black hole with constant surface gravity, the appropriate modes are well known.
They take the form $e^{in\kappa v}$, where $v$ is the advanced time along the horizon, 
normalized so that $\ell^a\nabla_av=1$.  Such modes are periodic in imaginary time with
period $2\pi/\kappa$, as required for nonsingular Greens functions.  
Here, though, $\kappa$ is one of our canonical variables, and we cannot simply take it to be 
constant.  In the language of \cite{Ashtekar}, we are considering ``nonexpanding
horizons'' but not ``isolated horizons.''   While we can always perform a local Lorentz
transformation to make $\kappa$ constant, that would require much more restrictive
boundary conditions, which would hide part of the symmetry.

Fortunately, though, the appropriate generalization is straightforward.  Define a phase 
$\psi$ such that
\begin{align}
D\psi \tq \kappa, \ {\bar D}\psi \tq 0 \quad\Leftrightarrow\ \ d\psi\tq -\kappa n_a
\quad \Leftrightarrow\ \ \psi \tq -\int_\Delta \kappa n_a \tq -\int\kappa dv \, .
\label{e3}
\end{align}
The modes are then
\begin{align}
\zeta_n \tq \frac{1}{\kappa}e^{in\psi} \qquad\hbox{(where $\zeta$ is either $\xi$ or $\eta$)} \, .
\label{e4}
\end{align}
The prefactor of $1/\kappa$ has been chosen so the modes obey the ordinary
algebra of diffeomorphisms of the circle,
\begin{align}
\{\zeta_m,\zeta_n\} = \zeta_mD\zeta_n - \zeta_nD\zeta_m = -i(m-n)\zeta_{m+n} \, .
\label{ex1}
\end{align}
Setting  $L_n = L[\xi_n]$ and $M_n = M[\eta_n]$, it is easy to check that our BMS${}_3$ 
algebra reduces to (\ref{xxe1}), with a central term
\begin{align}
-\frac{1}{16\pi G}\int_\Delta \left(D\xi_m D^2\eta_n - D\eta_n D^2\xi_m\right)n_a
   = -\frac{i}{16\pi G}\int_\Delta (mn^2-nm^2)e^{i(m+n)\psi}d\psi \, .
\label{e6}
\end{align}
If we take the integral to be over a single period---essentially mapping the problem
to a circle, as is standard in conformal field theory---we obtain a central charge 
\begin{align}
c_{\scriptscriptstyle LM} = \frac{1}{4G} \, .
\label{e7}
\end{align} 

We also need the zero-modes of  $L$ and $M$.  For $M$, this is straightforward: from (\ref{d7}),
\begin{align}
M_0  = M[\eta_0] = -\frac{1}{16\pi G}\int_\Delta \kappa^2\eta_0 n_a 
   = \frac{1}{16\pi G}\int d\psi = \frac{1}{8G} \, .
\label{e8}
\end{align}
For $L$, the ``bulk'' contribution to $L_0$ vanishes.  But $L$, unlike $M$, 
has a boundary contribution.  Indeed,  the variation leading to (\ref{d2}) involves integration 
by parts, with a boundary term
\begin{align}
\delta L[\xi] = \dots + \frac{1}{8\pi G}\left[\xi D\delta\varphi 
   - (D+\kappa)\xi\,\delta\varphi\right]\Bigl|_{\partial\Delta} \, .
\label{e9}
\end{align}
As noted in section \ref{secsympstruc}, the covariant phase space approach requires that we set 
$D\delta\varphi$ to zero at  the bifurcation point $B$.  We should certainly not hold $\varphi$ 
itself fixed, though, since that would fix $\varphi$ along the entire horizon, eliminating the 
shift symmetry.  Instead,  we should fix the conjugate variable $\kappa$ at $B$.  This 
requires an added boundary contribution to cancel the variation (\ref{e9}),
\begin{align}
L_0^{\hbox{\tiny bdry}} = \frac{1}{8\pi G} \varphi (D + \kappa)\xi_0 \,\Bigl|_{B} 
      = \frac{\varphi_+}{8\pi G} \, ,
\label{e10}
\end{align}
where $\varphi_+$ is the value of $\varphi$ at $B$.

\section{Entropy}

We are finally in a position to compute the entropy of our black hole.  Inserting (\ref{e7}), 
(\ref{e8}), and (\ref{e10}) into (\ref{xxe3}), we obtain  
\begin{align}
S = \frac{\varphi_+}{4G} \, .
\label{e11}
\end{align}

For a purely two-dimensional theory, this is the correct Bekenstein-Hawking entropy for
a black hole \cite{Kunstatter,Grumiller}.  From the $D$-dimensional perspective, it is
the contribution of a single null generator of the horizon.  But the symmetries that
determine (\ref{e11}) act independently on each generator, and entropy is an extensive
quantity, so we can add the individual entropies:
\begin{align}
S = \frac{1}{4G} \int d^{D-2}y\,\varphi_+ =  \frac{1}{4G} \int d^{D-2}y\,\sqrt{\phi_+} 
   = \frac{A_+}{4G} \, .
\label{e11xx}
\end{align}
where $A_+$ is the area of the bifurcation sphere.  We have thus obtained the correct
Bekenstein-Hawking entropy.for the full $D$-dimensional theory.

\section{Conclusions and directions}

As anticipated, black hole entropy is indeed determined by the symmetries
of the horizon.  In contrast to previous efforts to demonstrate this behavior, the   
derivation presented here has required no stretched horizon, no extra angular dependence,
and no other ad hoc ingredients.  The main assumption has merely been that the 
dimensionally reduced horizon obeys the ``boundary conditions'' of section \ref{sechor}.

What is the meaning of the crucial BMS${}_3$ symmetry?  It is not a gauge symmetry:
physical states are singlets under gauge symmetries, while our state-counting only works 
because the relevant states transform under high-dimensional representations.  This 
kind of behavior is typical of an asymptotic symmetry.  But our BMS${}_3$ is also
not quite a standard asymptotic symmetry: while we can view the horizon as a sort of 
boundary, it is a boundary that exists only for a restricted class of field configurations.  
Physically, we are asking a question of conditional probability---\emph{if} a black hole is 
present, what are its properties?---and the symmetries reflect this condition.  This is
at least vaguely analogous to entanglement entropy, which requires a similar specification 
of a boundary.  Indeed, it is possible that our horizon degrees of freedom might be viewed 
as a remnant left behind after tracing out the state behind the horizon.  For three-dimensional
topological field theory, this argument can be made fairly rigorous \cite{Witten};
it would be interesting to investigate it further in the present context.

There are several obvious directions for generalization.  A BMS symmetry at the horizon has
appeared in other settings (for instance, \cite{Donnay,Eling,Hawking,Fareghbal,Afshar,Donnayb}); 
the relationship to the BMS${}_3$ symmetry described here should be clarified.  Perhaps most 
fundamentally, if this symmetry really does explain the universality of black hole entropy, it should 
be present---although possibly hidden---in other derivations of entropy.  Hints of such a  hidden 
symmetry have been found for loop quantum gravity \cite{Carlip7},  induced gravity \cite{Frolov}, 
and near-extremal black holes in string theory \cite{Carlip8}, but none of these investigations has 
yet exploited the full BMS${}_3$ symmetry.  

Ideally, we might hope to do even more.  Many of the fundamental questions in black hole
thermodynamics involve the dynamics of Hawking radiation and its coupling to gravitational
degrees of freedom.  In 2+1 dimensions, Emparan and Sachs have succeeded in using the 
asymptotic conformal symmetry to couple the BTZ black hole to matter and obtain Hawking 
radiation \cite{Emparan}.  Perhaps our BMS${}_3$ symmetry will ultimately allow us to do the 
same in arbitrary dimensions.

\vspace{1.5ex}
\begin{flushleft}
\large\bf Acknowledgments
\end{flushleft}

This research was supported by the US Department of Energy under grant DE-FG02-91ER40674.

\appendix
\section{The covariant canonical formalism \label{secapp}}

The idea underlying the covariant canonical formalism is that for a theory with a well-posed initial 
value problem---that is, well-defined and unique time evolution---the phase space, viewed as the 
space of initial data, is isomorphic to the space of classical solutions \cite{AshMag,Bombelli,Waldb,Crn}.  
The isomorphism is not canonical, but requires a choice of a Cauchy surface $\Sigma$.  Once
$\Sigma$ has been chosen, though, the identification is simple: initial data on $\Sigma$
determines a unique classical solution, and a classical solution restricted to $\Sigma$
defines a unique set of initial data.  This equivalence, which can be traced back to Lagrange 
(see \cite{Bombelli}), means that we can formulate all the usual ingredients of Hamiltonian 
mechanics without ever having to break general covariance by choosing a particular time slicing.

Consider a theory in a $D$-dimensional spacetime with fields $\Phi^A$ (for us,
$\varphi$, $g$, and $\psi$) and a Lagrangian density $L[\Phi]$, which we view as a
$D$-form.  Under a general variation of the fields, $L[\Phi]$ changes as
\begin{align}
\delta L = E_A\delta\Phi^A + d\Theta[\Phi,\delta\Phi] \, ,
\label{a6}
\end{align}
where the equations of motion are $E_A=0$ and the last ``boundary'' term comes from 
integration by parts.  We normally ignore this boundary term, but in  the covariant 
canonical formalism it is crucial.  The symplectic current $\omega$ is 
defined by a second variation,
\begin{align}
\omega[\Phi;\delta_1\Phi,\delta_2\Phi] 
   = \delta_1\Theta[\Phi,\delta_2\Phi] - \delta_2\Theta[\Phi,\delta_1\Phi] \, ,
\label{ab6}
\end{align}
and the symplectic form is
\begin{align}
\Omega[\Phi;\delta_1\Phi,\delta_2\Phi] 
    = \int_\Sigma \omega[\Phi;\delta_1\Phi,\delta_2\Phi]
    = \int_\Sigma \omega_{AB}\delta_1\Phi^A\wedge\delta_2\Phi^B \, ,
\label{ab7}
\end{align}
where $\Sigma$ is a Cauchy surface.   (More precisely, $\Omega$ is often a presymplectic 
form, with degenerate directions that must be factored out to obtain a true symplectic form 
\cite{Waldb}.)

In keeping with the covariant phase space philosophy, $\Omega[\Phi;\delta_1\Phi,\delta_2\Phi]$ 
depends on a classical solution $\Phi$, which fixes a point in the phase space.  $\Omega$ 
itself is a two-form on the phase space, and the variations $\delta\Phi$ are tangent vectors 
to the space of classical solutions, that is, solutions of the linearized equations of motion.  For
a field theory in flat spacetime, it is not hard to check that when $\Sigma$ is a surface of 
constant time, (\ref{ab7}) is equivalent to the ordinary symplectic form. 

The symplectic current  (\ref{ab6}) is closed:
\begin{align}
d\omega[\Phi;\delta_1\Phi,\delta_2\Phi] 
   = \delta_1 d\Theta[\Phi,\delta_2\Phi] - \delta_2 d\Theta[\Phi,\delta_1\Phi]
   = -\delta_1 E_A\wedge \delta_2 \Phi^A + \delta_2 E_A\wedge \delta_1 \Phi^A = 0 \, ,
\label{ab7a}
\end{align}
since the variations satisfy the linearized equations of motion $\delta  E_A=0$.
Hence the symplectic form (\ref{ab7}) will depend only weakly on the choice of Cauchy 
surface: integrals over two surfaces $\Sigma_1$ and $\Sigma_2$ can differ only by 
boundary terms that might arise if $\partial\Sigma_1\ne\partial\Sigma_2$.   In particular,
for a diffeomorphism-invariant theory, a diffeomorphism generated by a vector field
$\zeta^a$ transverse to $\Sigma$ may be viewed as a deformation of the Cauchy surface,
and we have
\begin{align}
\Omega[\Phi;\delta_1\Phi,\delta_\zeta\Phi] = 0
\label{abxx}
\end{align}
as long as $\zeta^a$ vanishes at $\partial\Sigma$.  This may be checked explicitly
for the symplectic form (\ref{ac7}): under a diffeomorphism generated by a vector field 
$\zeta^a = {\bar\zeta} n^a$, one finds
\begin{align}
\Omega_\Delta[(\varphi,g);\delta(\varphi,g),\delta_\zeta(\varphi,g)] =
    \frac{1}{8\pi G}{\bar\zeta}\delta({\bar D}\varphi)\Bigl|_{\partial\Delta} + \dots \, ,
\label{ac10xx}
\end{align}
where the omitted terms are proportional to either the equations of motion or their first
variations, which are both set to zero in the covariant canonical formalism.  

As in ordinary mechanics, the symplectic form determines Poisson brackets 
and Hamiltonians.  Schematically, the Poisson bracket of two functions $X$ and $Y$ is
\begin{align}
\left\{X,Y\right\} 
   = \int_\Sigma\,\frac{\delta X}{\delta\Phi^A} (\omega^{-1})^{AB}\frac{\delta Y}{\delta\Phi^B} \, .
\label{ab10xx}
\end{align}
Given a family of transformations $\delta_\tau\Phi^A$ labeled by a parameter $\tau$, the 
Hamiltonian $H[\tau]$ that generates the transformations is determined by the condition
\begin{align}
\delta H[\tau] = \Omega[\delta\Phi,\delta_\tau\Phi]
\label{ab8}
\end{align}
for an arbitrary variation $\delta\Phi$.  Using (\ref{ab7}), we can see that this is just a 
disguised form of Hamilton's equations of motion,
\begin{align}
\delta_\tau\Phi^A = (\omega^{-1})^{AB}\,\frac{\delta H[\tau]}{\delta\Phi^B} \, .
\label{ab9b}
\end{align}
The Poisson bracket of two such generators is
\begin{align}
\left\{ H[\tau_1],H[\tau_2]\right\} = \delta_{\tau_2}H[\tau_1] 
  = -\Omega[\delta_{\tau_1}\Phi,\delta_{\tau_2}\Phi] \, .
\label{ab10}
\end{align}

\section{Gaussian null coordinates and dimensional reduction \label{AppB}}

In section \ref{secreduc}, Gaussian null coordinates were used to help reduce the
$D$-dimensional Einstein-Hilbert action to an effective two-dimensional form.  Here
I describe these coordinates in a bit more detail.

We start with the Gaussian null coordinate system\footnote{My notation differs a bit from 
that of \cite{Wald}: we use different index conventions,  my $v$  is their $u$, my $\ell^a$ 
is their $k^a$, and my $n_a$ is their $\ell_a$.}  described in Appendix A of \cite{Wald}
and in \cite{Booth}.  In such coordinates, a general metric takes the form
\begin{align}
ds^2 = -r\cdot fdv^2 + 2drdv + 2r\cdot h_\mu\,dv dy^\mu + \phi_{\mu\nu}dy^\mu dy^\nu \, .
\label{ABx1}
\end{align}
The surface $r=0$ is null; here we will take it to be the horizon $\Delta$
 
The coordinates (\ref{ABx1}) have clear geometrical meanings.  The horizon $r=0$ 
is a null surface with null normal $\ell_A dz^A = dr$.  (The surface gravity is 
$\kappa = \frac{1}{2}f|_{r=0}$).  Since $\Delta$ is null, its normals are also tangent 
vectors; indeed, the integral curves of the tangent vectors $\ell^A\partial_A 
= \frac{\partial\ }{\partial v}$ are the null geodesic generators of $\Delta$.  The ``orthogonal''
vectors $n^A\partial_A = \frac{\partial\ }{\partial r}$ are null even off the horizon,
and their integral curves are null geodesics transverse to the horizon.  As in section 
\ref{secreduc}, the coordinates $y^\mu$ parametrize a spacelike cross section 
$\hat\Delta$ of the horizon, and may be extended to a neighborhood of $\Delta$ by 
requiring that they be constant on both sets of null geodesics. 

The coordinate $r$ is an affine parameter along the transverse geodesics, and thus
provides a natural geometric notion of ``distance from the horizon.''  For the Schwarzschild 
metric, in particular, Gaussian null coordinates are Eddington-Finkelstein coordinates
with $r$ shifted to vanish at the horizon.  Near the horizon, $r\approx \rho^2/8m$, 
where $\rho$ is the proper distance to the horizon at constant time.  Appendix A of \cite{Booth}  
gives an explicit expression for the Kerr-Newman metric in Gaussian null coordinates.

As claimed in section \ref{secreduc}, the components $A_a{}^\mu$ of Yoon's metric 
(\ref{xy2})---here of the form $r\cdot h_\mu$---vanish on the horizon, and are $\mathcal{O}(r)$ 
near $\Delta$.   In fact, a direct calculation in these coordinates shows that the ordinary 
two-dimensional scalar curvature $R$ differs from the quantity $\hat R$ of eqn.\ (\ref{xy5}) 
by terms of order $r^2$, justifying the near-horizon form (\ref{a01}) of the action.  

The metric (\ref{ABx1}) is of the general Kaluza-Klein-like form (\ref{xy2}).   But Gaussian null 
coordinates are too restrictive  to exhibit the full set of available symmetries.  As in section 
\ref{secreduc}, though, we can move out of the Polyakov-like gauge by allowing an arbitrary 
two-dimensional coordinate transformation $x\rightarrow{\bar x}(x)$.  This will restore the 
general structure of the metric (\ref{xy2}),  while still restricting the $y$ dependence of the 
metric; for instance, although it will no longer be the case that $\sqrt{-g}=1$, it will remain true 
that $\partial_\mu\sqrt{-g} = 0$.  It may be checked that after such a transformation,  $R$ 
continues to differ from $\hat R$ only by terms of order $r^2$.

We may next ask how the $D$-dimensional vectors $\ell^A$ and $n^A$ are related to their
two-dimensional counterparts $\ell^a$ and $n^a$ of section \ref{secdil}.  For $\ell^A$, this is simple:
we have only defined $\ell^A$ on the horizon, where it is the tangent field to the null generators 
of $\Delta$, and thus coincides with $\ell^a$.  For $n^A$, the essential feature is that its
$D$-dimensional integral curves are affinely parametrized null geodesics:
\begin{align}
n^B\nabla_B n_A = 0 = n^B(\nabla_B n_A - \nabla_A n_B) = n^B(\partial_B n_A - \partial_A n_B) \, ,
\label{ABx2}
\end{align}
where I have used the fact that $n_An^A=0$.  Now, the only nonvanishing components of 
$n$ in Gaussian null coordinates  are $n^r$ and $n_v$, and under two-dimensional 
coordinate transformations $x\rightarrow{\bar x}(x)$ it remains true that only
two-dimensional components $n^a$ and $n_b$  are present.  Thus (\ref{ABx2}) becomes
\begin{align}
n^b(\partial_b n_a - \partial_a n_b) = 0 = n^b\,{}^{\scriptscriptstyle(2)}\nabla_bn_a \, ,
\label{ABx3}
\end{align}
which was the defining property of $n_a$ in section \ref{secdil}.  The $D$-dimensional 
transverse vectors $n_A$ thus coincide with the two-dimensional vectors $n_a$.  In particular,
the affine parameter $r$ gives a good measure of distance from the horizon in both $D$
and two dimensions.
 
Note that if $F$ is any function that vanishes 
at the horizon and is smooth near $\Delta$,
\begin{align}
F = r\partial_rF + \mathcal{O}(r^2) = r{\bar D}F + \mathcal{O}(r^2) \, .
\label{ABx4}
\end{align}
In particular,
\begin{align}
D\varphi = r{\bar D}D\varphi + \mathcal{O}(r^2) \, ,
\label{ABx4a}
\end{align}
quantifying the notion that  $D\varphi$ is a measure of distance from the horizon.

\section{Some details of near-horizon symmetries \label{AppC}}

Section \ref{secsym} discussed a ``near-horizon symmetry'' that played a crucial
role in counting states.  Specifically, I argued that for any $\eta$ satisfying ${\bar D}\eta\tq0$,
the transformation
\begin{subequations}
\begin{align}
&{\hat\delta}_\eta\varphi = \nabla_a(\eta\ell^a) + \mathcal{L}_\zeta \varphi \, , \label{AB1a} \\[.3ex]
&{\hat\delta}_\eta g_{ab} = X_\eta\frac{D\varphi}{{\bar D}D\varphi} g_{ab}  
    + \mathcal{L}_\zeta g_{ab}\label{AB1b} \, , \\[.3ex]
&{\hat\delta}_\eta \chi = \mathcal{L}_\zeta \chi \, ,\label{AB1c} \\[.3ex]
&\hbox{with}\ \
  \zeta^a = - \frac{D(D+\kappa)\eta}{{\bar D}D\varphi}n^a  \quad \hbox{and}\quad
  2D(D+\kappa)X_\eta + {\bar\zeta} {\bar D}DR = 0 
  \label{AB1d}
\end{align}
\end{subequations}
is an ``approximate symmetry'' of the action (\ref{a01}) with horizon boundary conditions 
(\ref{cxa})--(\ref{cxc}), in the sense that the variation of the action could be made ``arbitrarily
small.''   

As stated, this claim is a bit ambiguous.  First of all, the variation ${\hat\delta}_\eta I$ will be
inherently small as the support of $\eta$ shrinks to a small neighborhood of
the horizon, simply because the integration region becomes small.  Second, for this particular
variation it may be seen from (\ref{d11a})--(\ref{d11c}) that the symmetry algebra is unchanged
under a constant rescaling $\eta\rightarrow k\eta$, while the variation ${\hat\delta}_\eta I$ of
the action scales by $k$.  It is thus not entirely clear what ``small'' means.  

To remove these ambiguities, let us define an approximate near-horizon symmetry as one 
for which the quantity
\begin{align}
{\bar\delta}_\eta I = {\hat\delta}_\eta I\big/ {\textstyle \int\! |\eta|\epsilon}
\label{ABz1}
\end{align}
becomes arbitrarily small as the support of $\eta$ shrinks to a small enough neighborhood of 
the horizon.  (The absolute value in the denominator eliminates problems that could 
occur if $\int_\Delta \eta\, n_a = 0$.)  This expression is invariant under rescalings of $\eta$,
and the integral in the denominator compensates for the effects of a shrinking region of integration.
If, as in eqn.\ (\ref{AB6}) below, $\eta$ has support only in a band $r<\varepsilon$, this condition 
is roughly equivalent to normalizing $\eta$ at the horizon and then demanding that ${\hat\delta}_\eta I$
 go to zero faster than $\varepsilon$.

To apply this criterion to the transformations (\ref{AB1a})--(\ref{AB1c}), we should first
check that they preserve  our boundary conditions (\ref{cxa})--(\ref{cxc}).  Condition (\ref{cxc}) 
simply tells us that the Weyl transformation ${\hat\delta}_\eta g_{ab}$ 
acts only on $\ell_a$ and not on $n_a$ (and therefore on $n^a$ and not $\ell^a$).  Condition 
(\ref{cxa}) then determines the form of the transverse diffeomorphism $\bar\zeta$, and as show 
in section \ref{secsym}, condition (\ref{cxb}) gives the equation in (\ref{AB1d}) that fixes $X_\eta$.

We next examine the effect of this transformation on the action (\ref{a01}).  The
action is diffeomorphism invariant, so we can ignore $\zeta^a$ and consider only the shift 
of the dilaton and the Weyl transformation of the metric.  The variation of the action is then
\begin{align}
\delta_\eta I &= \int \left[\frac{\delta I}{\delta\varphi}{\hat\delta}_\eta\varphi 
       + \frac{\delta I}{\delta g_{ab}}{\hat\delta}_\eta g_{ab}\right]\epsilon \nonumber\\
       &= \frac{1}{16\pi G} \int \left[ \nabla_a(\eta\ell^a)\left(R + \frac{dV}{d\varphi}\right)
       + X_\eta\frac{D\varphi}{{\bar D}D\varphi}\left( -\Box\varphi + V\right)\right]\epsilon 
       = \frac{1}{16\pi G} \int\left[\eta A + X_\eta B\right]\epsilon 
\label{AB4}
\end{align} 
with
\begin{subequations}
\begin{align}
A &= -D\left(R + \frac{dV}{d\varphi}\right) = -DR - D\varphi\,\frac{d^2V}{d\varphi^2} 
      = -r\left( {\bar D}DR + {\bar D}D\varphi\,\frac{d^2V}{d\varphi^2}\right) + \mathcal{O}(r^2)
      \label{AB5a}\\
B& = \frac{D\varphi}{{\bar D}D\varphi}\left( -\Box\varphi + V\right) 
      = \frac{D\varphi}{{\bar D}D\varphi}\left( 2{\bar D}D\varphi + V\right) 
      = r\left( 2{\bar D}D\varphi + V\right)  + \mathcal{O}(r^2) \, ,
       \label{AB5b}
\end{align}
\end{subequations}
where in the last equalities I have used (\ref{ABx4}) to write the result in Gaussian null 
coordinates.

We have assumed that the parameter $\eta$ falls off rapidly away from the horizon---this is,
after all, a ``near-horizon'' symmetry.  Let us make this explicit by writing
\begin{align}
\eta = \eta_\Delta  \cdot \Psi_{\varepsilon}(r) \, ,
\label{AB6}
\end{align}
where $\eta_\Delta$ is the restriction of $\eta$ to the horizon and $\Psi$ is a smooth bump function
\begin{align}
&\Psi_{\varepsilon}(r) = \left\{\begin{array}{ll} 1 & r=0 \\ 
      \hbox{smooth interpolation}\quad  & 0<r<\varepsilon\\ 
      0 & r>\varepsilon \end{array}\right. \nonumber\\[1.4ex]
&\hbox{with} \quad\partial_r \Psi_{\varepsilon}\bigl|_{r=0} \,= 0 \, ,
\label{AB7}
\end{align}
where the last condition ensures that ${\bar D}\eta \tq 0$.  The variation (\ref{AB4})
is then
\begin{align}
\delta_\eta I = \frac{\varepsilon^2}{32\pi G} \int_\Delta \left[ 
     -\eta_\Delta\left( {\bar D}DR + {\bar D}D\varphi\frac{d^2V}{d\varphi^2}\right)
     + X_{\eta_\Delta}\left( 2{\bar D}D\varphi + V\right) \right]n_a + \mathcal{O}(\varepsilon^3) \, .
\label{AB8}
\end{align}
The denominator in (\ref{ABz1}), on the other hand, is
\begin{align}
\int |\eta|\epsilon = \varepsilon \int_\Delta |\eta_\Delta|n_a+ \mathcal{O}(\varepsilon^2) \, .
\label{AB9}
\end{align}
Thus as long as the integrand in (\ref{AB8}) remains well-behaved near the horizon,
the variation ${\bar\delta}_\eta I$ is of order $\varepsilon$, and can be made arbitrarily small
by shrinking the support of $\eta$.  

Note that while $\eta$ must have large ($\mathcal{O}(1/\varepsilon)$) radial derivatives, 
these never appear in the variation of the action.  For the first term in (\ref{AB8}) this is obvious; 
for the second, it follows from the fact that the defining equation (\ref{AB1d}) for $X_\eta$ involves 
no radial derivatives.  Recall also from section \ref{secreduc} that the corrections to the near-horizon 
form (\ref{a01}) of the action are at most of order $r$, so any additional variation of the
action coming from these terms will also fall off as $\varepsilon^2$.

\end{document}